# LOCALIZATION OF TWO-DIMENSIONAL FIVE-STATE QUANTUM WALKS

**CLEMENT AMPADU**

31 Carrolton Road
Boston, Massachusetts, 02132
USA
e-mail: drampadu@hotmail.com

## Abstract

We investigate a generalized Hadamard walk in two dimensions with five inner states. The particle governed by a five-state quantum walk (5QW) moves, in superposition, either leftward, rightward, upward, or downward according to the inner state. In addition to the four degrees of freedom, it is allowed to stay at the same position. We calculate rigorously the wave function of the particle starting from the origin in the plane for any initial state, and give the spatial distribution of probability of finding the particle. We also investigate the localization problem for the two-dimensional five-state quantum walk: Does the probability of finding a particle anywhere on the plane converge to zero even after infinite time steps except initial states?

**KEY WORDS:** localization, quantum walk, time-averaged probability, stationary distribution

**PACS:** 05.30.-d, 03.67.Lx, 05.40.-a, 89.70.+c

I. Introduction

The Hadamard walk as is well known plays a key role in the studies of the quantum walks, thus the generalization of the Hadamard walk is one of the many fascinating challenges. The simplest and well studied example of the Hadamard walk, Ref [1] for example, is given by the following unitary matrix $H = \frac{1}{\sqrt{2}}\begin{bmatrix} 1 & 1 \\ 1 & -1 \end{bmatrix}$. Some generalizations of this matrix include

$$H(\eta,\phi,\psi) = \frac{e^{i\eta}}{2}\begin{bmatrix} e^{i(\phi+\psi)} & e^{-i(\phi-\psi)} \\ e^{i(\phi-\psi)} & e^{-i(\phi+\psi)} \end{bmatrix} \text{ and } H^*(p,q) = \begin{bmatrix} \sqrt{p} & \sqrt{q} \\ \sqrt{q} & -\sqrt{p} \end{bmatrix}. \text{ In particular, } H^*(p,q) \text{ was}$$

considered by the author in a recent paper [2], as well as its two-dimensional counterpart

$H^{**}(p,q) = H^*(p,q) \otimes H^*(p,q)$. On the other hand $H(\eta,\phi,\psi)$ is well known by many authors including those in Ref [3]. In this paper we study a certain generalization of the Hadamard walk as described in the Abstract in two dimensions and clarify theoretically the following question: If, say, a quantum walker which could be a quantum particle exists only at one site initially in the plane, will the quantum walker remain trapped with high probability near the initial position? For the purposes of this paper we have reinterpreted this question which is given in the Abstract.

The answer to this question has been investigated by various authors in varying contexts under the pseudonym "localization". Inuit et.al [4] studied a generalized Hadamard walk in one dimension with three inner states and concluded that the quantum walker (quantum particle) is trapped near the origin with probability. Watabe et.al [5] were able to control localization around the origin for a one-parameter family of discrete-time quantum walk models on the square lattice, which included the Grover walk, which is related to the Grover's algorithm in computer science, as a special case. For the Grover walk itself in two dimensions, Inui et.al [6] were able to show localization analytically, whilst Mackay et.al [7] showed localization through simulations. It appears the first simulation showing localization is the work of Mackay et.al [7], after that more refined simulations were performed by Tregenna et.al [8]. Liu and Pentulante [9] were able to offer a theoretical explanation for localization in the case of discrete quantum random walks on a linear lattice with two entangled coins. Konno et.al [10] were able to generalize the criterion that the time averaged limit measure is positive for a class of one-dimensional quantum walks exhibiting localization, by focusing on the limiting theorems. For a certain quantum walk driven by two coins in which the conditional shift operator has three directions: forward, backward, or stall, Liu [11], was able to offer thorough analytic treatments of the asymptotic behaviors of the position probability distributions, and show that localization exist in certain instances depending on the initial state of the coin and the coin parameters. Schreiber et.al [12] investigated the impact of decoherence and static disorder on the dynamics of quantum particles moving in a periodic lattice, and were able to exhibit (Anderson) localization through simulations by applying controlled time-dependent operations. Ahlbrecht et.al [13] studied a spin-$\frac{1}{2}$ particle moving on a one dimensional lattice subject to disorder induced by a random, space-dependent quantum coin. The discrete time evolution is given by a family of unitary quantum walk operators, where the shift operation is assumed to be deterministic. Each coin is an independent identically distributed random variable with values in the group of two dimensional unitary matrices. The authors were able to derive sufficient conditions on the probability distribution of the coins exhibiting (dynamical) localization. Machida et.al [14] studied a certain 2-state quantum walk on the line defined by two matrices, were one of the matrices operates the walk at only half-time, and were able to show that the walk can be localized around the origin. Konno et.al [15] studied discrete-time quantum walks on the half-line by means of spectral analysis, and were able to rigorously justify localization of the quantum walk on homogeneous trees as shown by Chisaki et.al [16]. In [17], Chandrashekar presented an approach to induce localization of a Bose-Einstein condensate in a one-dimensional lattice under the influence of unitary quantum walk evolution using disordered quantum coin operation. Kollar et.al [18], studied a three-state quantum walk on triangular lattices associated with the Grover transformation, and showed that the three-state Grover walk does not exhibit localization, in sharp contrast to the Grover walk on the two dimensional square lattice. Shikano et.al [19] studied a class of discrete-time quantum walks on one -dimensional lattices in which the coin operator is position-dependent, these sort of quantum walks are called Inhomogeneous quantum walks in the literature. When the period of coin operators is incommensurable with respect to the lattice spacing, the authors show rigorously localization about the origin. Joye et.al [20] also studied the Inhomogeneous quantum walk in one dimension. In their case the dynamics of the walk is determined by repeated action of a coin operator in $U(2)$ on the internal degrees of freedom followed by a one step shift to the right or left, conditioned on the state the coin. When the coin operator depends on the position of the walker and is given by certain i.i.d. random process, the walk exhibits (Anderson) localization which is dynamic in nature. In the case of the space- inhomogeneous quantum walk in one dimension considered by Konno et.al [21] , show that the walk exhibits localization by a path counting method. For a one-parameter family of discrete-time quantum walk models in one and two dimensions associated with the Hadamard walk , theoretical criterion was given for the existence of localization in Ampadu [2]. A similar criterion was given in [22] for a quantum walk involving $M-$ particles.

As one can see from the literature review it is possible to exhibit localization of quantum walks experimentally, or to show localization occurs theoretically. Obviously the theoretical justification of localization is superior to the experimental realization as we argue in this paper. Therefore the aim of this paper is show that localization occurs for the discrete-time 5QW rigorously.

The rest of this paper is organized as follows. After defining the 5QW, the eigenvalues are calculated explicitly, we describe obtaining the eigenvectors via the Gram-Schmidt algorithm, and show the wave function in Sec.II. In Sec.III we argue why the experimental realization of localization of the 5QW via the time-averaged probability is not rigorous enough in answering the localization question. In Sec.IV we prove rigorously the localization of the 5QW by way of Theorem IV.1. Sec.V is devoted to the conclusions.

## II. Definition of the Five-State Quantum Walk

The five-state quantum walk is a kind of generalized Hadamard walk in the plane, and differs markedly from the previous studies. The particle ruled by the 5QW is characterized in the Hilbert space which is defined by a direct product of a chirality-state space

$|s\rangle \in \{|L\rangle, |R\rangle, |0\rangle, |D\rangle, |U\rangle\}$ and a position space spanned by $\{|n_1, n_2\rangle : n_1, n_2 \in Z\}$. The chirality states are transformed at each time step by the following unitary transformation:

$$|L\rangle = \frac{1}{5}\{-3|L\rangle + 2|R\rangle + 2|0\rangle + 2|D\rangle + 2|U\rangle\}, \quad |R\rangle = \frac{1}{5}\{2|L\rangle - 3|R\rangle + 2|0\rangle + 2|D\rangle + 2|U\rangle\},$$

$$|0\rangle = \frac{1}{5}\{2|L\rangle + 2|R\rangle - 3|0\rangle + 2|D\rangle + 2|U\rangle\}, \quad |D\rangle = \frac{1}{5}\{2|L\rangle + 2|R\rangle + 2|0\rangle - 3|D\rangle + 2|U\rangle\},$$

$$|U\rangle = \frac{1}{5}\{2|L\rangle + 2|R\rangle + 2|0\rangle + 2|D\rangle - 3|U\rangle\}.$$ Let

$$\Psi(n_1, n_2, t) \equiv [\psi_L(n_1, n_2, t) \quad \psi_R(n_1, n_2, t) \quad \psi_0(n_1, n_2, t) \quad \psi_D(n_1, n_2, t) \quad \psi_U(n_1, n_2, t)]^T \equiv [\alpha \quad \beta \quad \gamma \quad \lambda \quad \mu]$$

be the amplitude of the wave function of the particle corresponding to the chiralities $L, R, 0, D, U$ at the position $(n_1, n_2) \in Z^2$ and the time $t \in \{0, 1, 2, \cdots\}$, where $T$ denotes the transposition, and $Z$ is the set of integers. We assume that a particle exists initially at the origin in the plane, then the initial quantum states are determined by

$$\Psi(0,0,0) \equiv [\psi_L(0,0,0) \quad \psi_R(0,0,0) \quad \psi_0(0,0,0) \quad \psi_D(0,0,0) \quad \psi_U(0,0,0)]^T \equiv [\alpha \quad \beta \quad \gamma \quad \lambda \quad \mu], \text{ where}$$

$\alpha, \beta, \gamma, \lambda, \mu \in C$, and $C$ is the set of complex numbers. Before we define the time evolution of the wave function, we introduce the following operators:

$$U_L = \frac{1}{5}\begin{bmatrix} -3 & 2 & 2 & 2 & 2 \\ 0 & 0 & 0 & 0 & 0 \\ 0 & 0 & 0 & 0 & 0 \\ 0 & 0 & 0 & 0 & 0 \\ 0 & 0 & 0 & 0 & 0 \end{bmatrix}, \quad U_R = \frac{1}{5}\begin{bmatrix} 0 & 0 & 0 & 0 & 0 \\ 2 & -3 & 2 & 2 & 2 \\ 0 & 0 & 0 & 0 & 0 \\ 0 & 0 & 0 & 0 & 0 \\ 0 & 0 & 0 & 0 & 0 \end{bmatrix}, \quad U_0 = \frac{1}{5}\begin{bmatrix} 0 & 0 & 0 & 0 & 0 \\ 0 & 0 & 0 & 0 & 0 \\ 2 & 2 & -3 & 2 & 2 \\ 0 & 0 & 0 & 0 & 0 \\ 0 & 0 & 0 & 0 & 0 \end{bmatrix}$$

$$U_D = \frac{1}{5}\begin{bmatrix} 0 & 0 & 0 & 0 & 0 \\ 0 & 0 & 0 & 0 & 0 \\ 0 & 0 & 0 & 0 & 0 \\ 2 & 2 & 2 & -3 & 0 \\ 0 & 0 & 0 & 0 & 0 \end{bmatrix}, U_U = \frac{1}{5}\begin{bmatrix} 0 & 0 & 0 & 0 & 0 \\ 0 & 0 & 0 & 0 & 0 \\ 0 & 0 & 0 & 0 & 0 \\ 0 & 0 & 0 & 0 & 0 \\ 2 & 2 & 2 & 2 & -3 \end{bmatrix}.$$

Note that if the matrix, say $U_R$ is applied to $\Psi(n_1,n_2,t)$, then only the $R$ component is selected after carrying out the superposition between $\psi_L(n_1,n_2,t)$, $\psi_0(n_1,n_2,t)$, $\psi_D(n_1,n_2,t)$, $\psi_U(n_1,n_2,t)$. This is also similar for the other matrices $U_0$, $U_D$, $U_L$, $U_U$. We now define the time evolution of the wave function and this is given by

$$\Psi(n_1,n_2,t+1) = U_L\Psi(n_1+1,n_2,t) + U_R\Psi(n_1-1,n_2,t) + U_0\Psi(n_1,n_2,t) + U_D\Psi(n_1,n_2+1,t) + U_U\Psi(n_1,n_2-1,t)$$

One finds clearly that the chiralities $L, R, 0.D, U$ correspond to the left, right, neutral, downward, and upward state for the motion. Using the Fourier Analysis we can get the wave function. Notice that the spatial Fourier transform of $\Psi(k_{n_1},k_{n_2},t)$ is defined by

$$\tilde{\Psi}(k_{n_1},k_{n_2},t) = \sum_{(n_1,n_2)\in Z^2} \Psi(n_1,n_2,t)e^{-i(k_{n_1}n_1 + k_{n_2}n_2)}.$$ In the Fourier domain, the dynamics of the wave function is defined by $\tilde{\Psi}(k_{n_1},k_{n_2},t+1) = \tilde{U}\tilde{\Psi}(k_{n_1},k_{n_2},t)$, where

$$\tilde{U} = \frac{1}{5}\begin{bmatrix} e^{ik_{n_1}} & 0 & 0 & 0 & 0 \\ 0 & e^{-ik_{n_1}} & 0 & 0 & 0 \\ 0 & 0 & 1 & 0 & 0 \\ 0 & 0 & 0 & e^{ik_{n_2}} & 0 \\ 0 & 0 & 0 & 0 & e^{-ik_{n_2}} \end{bmatrix}\begin{bmatrix} -3 & 2 & 2 & 2 & 2 \\ 2 & -3 & 2 & 2 & 2 \\ 2 & 2 & -3 & 2 & 2 \\ 2 & 2 & 2 & -3 & 2 \\ 2 & 2 & 2 & 2 & -3 \end{bmatrix}.$$ The standard argument by induction on the time step allows us to write $\tilde{\Psi}(k_{n_1},k_{n_2},t) = \tilde{U}^t\tilde{\Psi}(k_{n_1},k_{n_2},0)$. We suppose that $\left|\Phi^j_{k_{n_1},k_{n_2}}\right\rangle$ is the orthonormal eigenvector of $\tilde{U}$ corresponding to the eigenvalue $e^{i\theta_{j,k_{n_1},k_{n_2}}}$ for $j = 1,2,3,4,5$. Since $\tilde{U}$ is unitary, it can be diagonalized. In particular, the wave function

$\tilde{\Psi}(k_{n_1}, k_{n_2}, t)$ can be expressed by $\tilde{\Psi}(k_{n_1}, k_{n_2}, t) = \left( \sum_{j=1}^{5} e^{i\theta_{j,k_{n_1},k_{n_2}}} |\Phi^j_{k_{n_1},k_{n_2}}\rangle\langle\Phi^j_{k_{n_1},k_{n_2}}| \right) \tilde{\Psi}(k_{n_1}, k_{n_2}, 0)$

where $\tilde{\Psi}(k_{n_1}, k_{n_2}, 0) = [\alpha \ \ \beta \ \ \gamma \ \ \lambda \ \ \mu]^T \in C^5$ with $|\alpha|^2 + |\beta|^2 + |\gamma|^2 + |\lambda|^2 + |\mu|^2 = 1$.

To be explicit we need to specify $\theta_{j,k_{n_1},k_{n_2}}$ for the eigenvalues of $\tilde{U}$ and these are given by

$$\theta_{j,k_{n_1},k_{n_2}} = \begin{cases} 0, & j=1 \\ -\theta^{-,-}_{k_{n_1},k_{n_2}}, & j=2 \\ -\theta^{+,-}_{k_{n_1},k_{n_2}}, & j=3 \\ \theta^{-,+}_{k_{n_1},k_{n_2}}, & j=4 \\ -\theta^{-,+}_{k_{n_1},k_{n_2}}, & j=5 \end{cases}$$

where $\sin\theta^{\pm,\pm}_{k_{n_1},k_{n_2}} = \dfrac{\sqrt{C(k_{n_1},k_{n_2})}}{10\sqrt{2}}$

$$\cos\theta^{\pm,\pm}_{k_{n_1},k_{n_2}} = -\frac{1}{10}A(k_{n_1},k_{n_2}) \pm \frac{\frac{1}{2}\left[\left(\frac{2}{25}A^2(k_{n_1},k_{n_2}) - \frac{2}{5}B(k_{n_1},k_{n_2}) \pm \left(\frac{-8}{125}A(k_{n_1},k_{n_2}) + 5\left[\frac{16}{25}A(k_{n_1},k_{n_2})B(k_{n_1},k_{n_2}) - \frac{16}{5}A(k_{n_1},k_{n_2})\right]\right)\right)\right]}{2\sqrt{2}\left(D(k_{n_1},k_{n_2})\right)^{\frac{1}{2}} - 2}$$

and

$A(k_{n_1}, k_{n_2}) = 3\cos(k_{n_1}) + 3\cos(k_{n_2}) + 4$

$B(k_{n_1}, k_{n_2}) = \cos(k_{n_1} - k_{n_2}) + \cos(k_{n_1} + k_{n_2}) + 4\cos k_{n_1} + 4\cos k_{n_2} + 5$

$C(k_{n_1}, k_{n_2}) = 2\cos(k_{n_1} - k_{n_2}) + 2\cos(k_{n_1} + k_{n_2}) + 32\cos(k_{n_1}) - 9\cos(2k_{n_1}) + 32\cos(k_{n_2}) - 9\cos(2k_{n_2}) - 50$

$D(k_{n_1}, k_{n_2}) = -2\cos(k_{n_1} - k_{n_2}) - 2\cos(k_{n_1} + k_{n_2}) - 32\cos(k_{n_1}) + 9\cos(2k_{n_1}) - 32\cos(k_{n_2}) + 9\cos(2k_{n_2}) + 50$

Remark that $\sin \theta_{k_{n_1},k_{n_2}}^{\pm,\pm}$ is independent of the superscript, but is convenient for the purposes of making the notation uniform. To obtain the orthonormal eigenvectors of $\tilde{U}$, we describe the procedure, the Gram-Schmidt algorithm. We consider a subspace spanned by the eigenvectors of $\tilde{U}$ and construct the orthonormal vectors $\left| \Phi_{k_{n_1},k_{n_2}}^{j} \right\rangle$ of the subspace spanned by the eigenvectors of $\tilde{U}$ as follows: Let

$$\left| \Phi_{k_{n_1},k_{n_2}}^{1} \right\rangle = \frac{1}{\left\| \tilde{\Phi}_{k_{n_1},k_{n_2}}^{1} \right\|} \left| \tilde{\Phi}_{k_{n_1},k_{n_2}}^{1} \right\rangle.$$

As we define $\left| \Phi_{k_{n_1},k_{n_2}}^{j} \right\rangle$ for $j = 2,\cdots,5$, we may assume that an orthonormal basis $\left| \Phi_{k_{n_1},k_{n_2}}^{1} \right\rangle, \cdots, \left| \Phi_{k_{n_1},k_{n_2}}^{j-1} \right\rangle$ of the subspace spanned by $\left| \tilde{\Phi}_{k_{n_1},k_{n_2}}^{1} \right\rangle, \cdots, \left| \tilde{\Phi}_{k_{n_1},k_{n_2}}^{j-1} \right\rangle$ has already been found. Let

$$\left| \Phi_{k_{n_1},k_{n_2}}^{j} \right\rangle = \frac{1}{\left\| \tilde{\Phi}_{k_{n_1},k_{n_2}}^{j} \right\rangle - proj_{\tilde{\Phi}_{k_{n_1},k_{n_2}}^{j-1}} \left| \tilde{\Phi}_{k_{n_1},k_{n_2}}^{j} \right\rangle \right\|} \left( \left| \tilde{\Phi}_{k_{n_1},k_{n_2}}^{j} \right\rangle - proj_{\tilde{\Phi}_{k_{n_1},k_{n_2}}^{j-1}} \left| \tilde{\Phi}_{k_{n_1},k_{n_2}}^{j} \right\rangle \right)$$

where

$$\left| \tilde{\Phi}_{k_{n_1},k_{n_2}}^{j} \right\rangle - proj_{\tilde{\Phi}_{k_{n_1},k_{n_2}}^{j-1}} \left| \tilde{\Phi}_{k_{n_1},k_{n_2}}^{j} \right\rangle = \left| \tilde{\Phi}_{k_{n_1},k_{n_2}}^{j} \right\rangle - \left( \left| \Phi_{k_{n_1},k_{n_2}}^{1} \right\rangle \cdot \left| \tilde{\Phi}_{k_{n_1},k_{n_2}}^{j} \right\rangle \right) \left| \Phi_{k_{n_1},k_{n_2}}^{1} \right\rangle - \cdots - \left( \left| \Phi_{k_{n_1},k_{n_2}}^{j-1} \right\rangle \cdot \left| \tilde{\Phi}_{k_{n_1},k_{n_2}}^{j} \right\rangle \right) \left| \Phi_{k_{n_1},k_{n_2}}^{j-1} \right\rangle$$

Note that the dot, $\cdot$, in the formulas above represent the dot product of two vectors. For example, using the Gram-Schmidt procedure we obtain the first orthonormal eigenvector of $\tilde{U}$ as follows

$$\left| \Phi_{k_{n_1},k_{n_2}}^{1} \right\rangle = \frac{1}{\left\| \tilde{\Phi}_{k_{n_1},k_{n_2}}^{1} \right\|} \left| \tilde{\Phi}_{k_{n_1},k_{n_2}}^{1} \right\rangle, \text{ where}$$

$$\left|\tilde{\Phi}^1_{k_{n_1},k_{n_2}}\right\rangle = \begin{bmatrix} -\dfrac{2\cos k_{n_2} + \sin^2 k_{n_2} + 1}{\cos k_{n_2} - \sin k_{n_2} + 1} - \dfrac{i(\sin k_{n_2} + \sin k_{n_1}) - \cos k_{n_1} - \cos k_{n_2} - 2}{\cos k_{n_2} - \sin k_{n_1} \sin k_{n_2} + \cos k_{n_1} + 1} \\ \\ \dfrac{\cos k_{n_1} - \sin k_{n_1} \sin k_{n_b} + \cos k_{n_2} + 1}{\cos k_{n_2} + \sin k_{n_1} \sin k_{n_2} + \cos k_{n_1} + 1} \\ \\ \dfrac{-0.5 i \sin k_{n_2} + \cos k_{n_2} + 1}{\cos k_{n_2} - i \sin k_{n_2}} \\ \\ \dfrac{2\cos k_{n_2} + \sin^2 k_{n_2} + 1}{\cos k_{n_2} - \sin k_{n_2} + 1} \\ \\ 1 \end{bmatrix}$$

Note that $\left\|\tilde{\Phi}^1_{k_{n_1},k_{n_2}}\right\rangle\right\|$ is the square root of the sum of squares of the entries in $\left|\tilde{\Phi}^1_{k_{n_1},k_{n_2}}\right\rangle$ given immediately above. For $j = 2,\cdots,5$ the remaining orthonormal eigenvectors can be calculated using the formulas

$$\left|\Phi^j_{k_{n_1},k_{n_2}}\right\rangle = \dfrac{1}{\left\||\tilde{\Phi}^j_{k_{n_1},k_{n_2}}\rangle - proj_{\tilde{\Phi}^{j-1}_{k_{n_1},k_{n_2}}} |\tilde{\Phi}^j_{k_{n_1},k_{n_2}}\rangle\right\|} \left(\left|\tilde{\Phi}^j_{k_{n_1},k_{n_2}}\right\rangle - proj_{\tilde{\Phi}^{j-1}_{k_{n_1},k_{n_2}}} \left|\tilde{\Phi}^j_{k_{n_1},k_{n_2}}\right\rangle\right)$$

where

$$\left|\tilde{\Phi}^j_{k_{n_1},k_{n_2}}\right\rangle - proj_{\tilde{\Phi}^{j-1}_{k_{n_1},k_{n_2}}} \left|\tilde{\Phi}^j_{k_{n_1},k_{n_2}}\right\rangle = \left|\tilde{\Phi}^j_{k_{n_1},k_{n_2}}\right\rangle - \left(\left|\Phi^1_{k_{n_1},k_{n_2}}\right\rangle \cdot \left|\tilde{\Phi}^j_{k_{n_1},k_{n_2}}\right\rangle\right) \left|\Phi^1_{k_{n_1},k_{n_2}}\right\rangle - \cdots - \left(\left|\Phi^{j-1}_{k_{n_1},k_{n_2}}\right\rangle \cdot \left|\tilde{\Phi}^j_{k_{n_1},k_{n_2}}\right\rangle\right) \left|\Phi^{j-1}_{k_{n_1},k_{n_2}}\right\rangle$$

Note that the dot, $\cdot$, in the formulas above represent the dot product of two vectors, and

$\left\||\tilde{\Phi}^j_{k_{n_1},k_{n_2}}\rangle - proj_{\tilde{\Phi}^{j-1}_{k_{n_1},k_{n_2}}} |\tilde{\Phi}^j_{k_{n_1},k_{n_2}}\rangle\right\|$ is the square root of the sum of squares of the entries in

$\left|\tilde{\Phi}^j_{k_{n_1},k_{n_2}}\right\rangle - proj_{\tilde{\Phi}^{j-1}_{k_{n_1},k_{n_2}}} \left|\tilde{\Phi}^j_{k_{n_1},k_{n_2}}\right\rangle$, which is defined above. In the above calculations, it is important to

note that the eigenvalue 1 is independent of the values $k_{n_1}, k_{n_2}$. In the one-dimensional setting, corresponding to the three-state quantum walk [4], the authors have shown that the eigenvalues of their generalized Hadamard walk are given by $e^{i\theta_{j,k}}$ where

$$\theta_{j,k} = \begin{cases} 0, j=1 \\ \theta_k, j=2 \\ -\theta_k, j=3 \end{cases}, \text{ where } \cos\theta_k = \frac{-1}{3}(2+\cos k) \text{ and } \sin\theta_k = \frac{1}{3}\sqrt{(5+\cos k)(1-\cos k)}. \text{ It is also}$$

seen that the eigenvalue 1 is independent of the value $k$. In particular, in [6] it is shown that existence of a strongly degenerate eigenvalue such as 1 as we have obtained in this paper is a necessary condition for localization. Number the chirality $L, R, 0, D, U$ using $l = 1,2,3,4,5$ respectively. The wave function in real space is obtained by the inverse Fourier transform for $\alpha, \beta, \gamma, \lambda, \mu \in C$ with

$$|\alpha|^2 + |\beta|^2 + |\gamma|^2 + |\lambda|^2 + |\mu|^2 = 1 \text{ as,}$$

$$\Psi(n_1,n_2,t;\alpha,\beta,\gamma,\lambda,\mu) = \frac{1}{4\pi^2}\int_{-\pi}^{\pi}\int_{-\pi}^{\pi}\tilde{\Psi}(k_{n_1},k_{n_2},t)e^{i(k_{n_1}n_1+k_{n_2}n_2)}dk_{n_1}dk_{n_2}$$

$$= \frac{1}{4\pi^2}\int_{-\pi}^{\pi}\int_{-\pi}^{\pi}\left(\sum_{j=1}^{5}e^{i\theta_{j,k_{n_1},k_{n_2}}t}|\Phi^j_{k_{n_1},k_{n_2}}\rangle\langle\Phi^j_{k_{n_1},k_{n_2}}|\tilde{\Psi}(k_{n_1},k_{n_2},0)\right)e^{i(k_{n_1}n_1+k_{n_2}n_2)}dk_{n_1}dk_{n_2}$$

$$= [\Psi(n_1,n_2,t;1;\alpha,\beta,\gamma,\lambda,\mu) \quad \Psi(n_1,n_2,t;2;\alpha,\beta,\gamma,\lambda,\mu) \quad \Psi(n_1,n_2,t;3;\alpha,\beta,\gamma,\lambda,\mu) \quad \Psi(n_1,n_2,t;4;\alpha,\beta,\gamma,\lambda,\mu) \quad \Psi(n_1,n_2,t;5;\alpha,\beta,\gamma,\lambda,\mu)]^T$$

$$= \sum_{j=1}^{5}[\Psi_j(n_1,n_2,t;1;\alpha,\beta,\gamma,\lambda,\mu) \quad \Psi_j(n_1,n_2,t;2;\alpha,\beta,\gamma,\lambda,\mu) \quad \Psi_j(n_1,n_2,t;3;\alpha,\beta,\gamma,\lambda,\mu) \quad \Psi_j(n_1,n_2,t;4;\alpha,\beta,\gamma,\lambda,\mu) \quad \Psi_j(n_1,n_2,t;5;\alpha,\beta,\gamma,\lambda,\mu)]^T$$

where $\Psi_j(n_1,n_2,t;l;\alpha,\beta,\gamma,\lambda,\mu) = \frac{1}{4\pi^2}\int_{-\pi}^{\pi}\int_{-\pi}^{\pi}c(\theta_{j,k_{n_1},k_{n_2}})\varphi_{k_{n_1},k_{n_2}}(\theta_{j,k_{n_1},k_{n_2}}l_j)e^{i(\theta_{j,k_{n_1},k_{n_2}}t+k_{n_1}n_1+k_{n_2}n_2)}dk_{n_1}dk_{n_2}$

with

$$\varphi_{k_{n_1},k_{n_2}}(\theta,l_j) = \zeta_{l_j,k_{n_1},k_{n_2}}(\theta)\{\alpha\overline{\zeta_{1,k_{n_1},k_{n_2}}(\theta)} + \beta\overline{\zeta_{2,k_{n_1},k_{n_2}}(\theta)} + \gamma\overline{\zeta_{3,k_{n_1},k_{n_2}}(\theta)} + \lambda\overline{\zeta_{4,k_{n_1},k_{n_2}}(\theta)} + \mu\overline{\zeta_{5,k_{n_1},k_{n_2}}(\theta)}\}$$

where $\zeta_{l_j,k_{n_1},k_{n_2}}(\theta)$ is the $l$ th entry of $|\Phi^j_{k_{n_1},k_{n_2}}\rangle$, $c(\theta_{j,k_{n_1},k_{n_2}})$ is the square of an appropriate normalization constant corresponding to $\left|\Phi^{j}_{k_{n_1},k_{n_2}}\right\rangle$, and the bar denotes complex conjugation. In this paper we shall sometimes write $\Psi_j(n_1,n_2,t;l;\alpha,\beta,\gamma,\lambda,\mu) = \Psi_j(n_1,n_2,t;l)$ when convenient.

The probability of finding the particle at position $(n_1,n_2)$ at time $t$ with chirality $l$ is given by

$P(n_1,n_2,t;l) = |\Psi(n_1,n_2,t;l)|^2$. Thus the probability of finding the particle at position $(n_1,n_2)$ at time $t$ is given by $P(n_1,n_2,t) = \sum_{l=1}^{5} P(n_1,n_2,t;l)$.

### III. Time-Averaged Probability

In this section we discuss some issues surrounding the experimental realization of the localization problem for the 5QW which sets the stage for the settling the localization problem theoretically. The time-averaged probability of the 5QW is defined by

$$\overline{P}_\infty(n_1,n_2,\alpha,\beta,\gamma,\lambda,\mu) = \lim_{N\to\infty}\left(\lim_{T\to\infty}\frac{1}{T}\sum_{l=1}^{5}\sum_{t=0}^{T-1}P_N(n_1,n_2,t;l;\alpha,\beta,\gamma,\lambda,\mu)\right), \text{ where}$$

$P_N(n_1,n_2,t;l;\alpha,\beta,\gamma,\lambda,\mu)$ is the probability of finding a particle with chirality $l$ at position $(n_1,n_2)$ at time $t$ in the plane containing $N$ sites. From the experimental view point, if localization exist in the plane we expect the time-averaged probability to converge to a non-zero value. However, we can only show that the time-averaged probability of the 5QW converge to a nonzero value except in special initial states, but this is not sufficient criterion for localization. In particular, it does not gurantee that the particle will be observed with almost the same probability after a long-time evolution anywhere in the plane. Particular examples include the four-state quantum walk [23,24] and the well known Grover walk which have shown to exhibit localization, but the probability of finding the particle oscillate. To answer the localization problem for the 5QW with rigor we look at the following probability in the next section $P(n_1,n_2) \equiv \lim_{t\to\infty} P(n_1,n_2,t)$, where $P(n_1,n_2,t) = P(n_1,n_2,t;\alpha,\beta,\gamma,\lambda,\mu)$. The idea of using the stationary probability distribution to settle the localization problem is not new and has been considered by a number of authors including Ampadu [2,23].

## IV. Stationary Probability Distribution of the Particle

Recall that the wave function is given by

$$\Psi(n_1,n_2,t;\alpha,\beta,\gamma,\lambda,\mu) = \frac{1}{4\pi^2}\int_{-\pi}^{\pi}\int_{-\pi}^{\pi}\tilde{\Psi}(k_{n_1},k_{n_2},t)e^{i(k_{n_1}n_1+k_{n_2}n_2)}dk_{n_1}dk_{n_2}$$

$$= \frac{1}{4\pi^2}\int_{-\pi}^{\pi}\int_{-\pi}^{\pi}\left(\sum_{j=1}^{5}e^{i\theta_{j,k_{n_1},k_{n_2}}t}\left|\Phi_{k_{n_1},k_{n_2}}^{j}\right\rangle\left\langle\Phi_{k_{n_1},k_{n_2}}^{j}\right|\tilde{\Psi}(k_{n_1},k_{n_2},0)\right)e^{i(k_{n_1}n_1+k_{n_2}n_2)}dk_{n_1}dk_{n_2}$$

$$= [\Psi(n_1,n_2,t;1;\alpha,\beta,\gamma,\lambda,\mu) \quad \Psi(n_1,n_2,t;2;\alpha,\beta,\gamma,\lambda,\mu) \quad \Psi(n_1,n_2,t;3;\alpha,\beta,\gamma,\lambda,\mu) \quad \Psi(n_1,n_2,t;4;\alpha,\beta,\gamma,\lambda,\mu) \quad \Psi(n_1,n_2,t;5;\alpha,\beta,\gamma,\lambda,\mu)]^T$$

$$= \sum_{j=1}^{5}[\Psi_j(n_1,n_2,t;1;\alpha,\beta,\gamma,\lambda,\mu) \quad \Psi_j(n_1,n_2,t;2;\alpha,\beta,\gamma,\lambda,\mu) \quad \Psi_j(n_1,n_2,t;3;\alpha,\beta,\gamma,\lambda,\mu) \quad \Psi_j(n_1,n_2,t;4;\alpha,\beta,\gamma,\lambda,\mu) \quad \Psi_j(n_1,n_2,t;5;\alpha,\beta,\gamma,\lambda,\mu)]^T$$

where $\Psi_j(n_1,n_2,t;l;\alpha,\beta,\gamma,\lambda,\mu) = \frac{1}{4\pi^2}\int_{-\pi}^{\pi}\int_{-\pi}^{\pi}c(\theta_{j,k_{n_1},k_{n_2}})\varphi_{k_{n_1},k_{n_2}}(\theta_{j,k_{n_1},k_{n_2}},l_j)e^{i\left(\theta_{j,k_{n_1},k_{n_2}}t+k_{n_1}n_1+k_{n_2}n_2\right)}dk_{n_1}dk_{n_2}$

with

$$\varphi_{k_{n_1},k_{n_2}}(\theta,l_j) = \zeta_{l_j,k_{n_1},k_{n_2}}(\theta)\left\{\alpha\overline{\zeta_{1_j,k_{n_1},k_{n_2}}(\theta)}+\beta\overline{\zeta_{2_j,k_{n_1},k_{n_2}}(\theta)}+\gamma\overline{\zeta_{3_j,k_{n_1},k_{n_2}}(\theta)}+\lambda\overline{\zeta_{4_j,k_{n_1},k_{n_2}}(\theta)}+\mu\overline{\zeta_{5_j,k_{n_1},k_{n_2}}(\theta)}\right\}$$

where $\zeta_{l_j,k_{n_1},k_{n_2}}(\theta)$ is the $l$ th entry of $\left|\Phi_{k_{n_1},k_{n_2}}^{j}\right\rangle$, $c(\theta_{j,k_{n_1},k_{n_2}})$ is the square of an appropriate normalization constant corresponding to $\left|\Phi_{k_{n_1},k_{n_2}}^{j}\right\rangle$, and the bar denotes complex conjugation. It is seen that the wave function is a linear combination of $e^{i\left(\theta_{j,k_{n_1},k_{n_2}}t+k_{n_1}n_1+k_{n_2}n_2\right)}$. If the argument $\theta_{j,k_{n_1},k_{n_2}}$ is not absolutely zero, then it is expected that $e^{i\left(\theta_{j,k_{n_1},k_{n_2}}t+k_{n_1}n_1+k_{n_2}n_2\right)}$ oscillates with high frequency for $k_x,k_y \in (-\pi,\pi]$ and large $t$, and $c(\theta_{j,k_{n_1},k_{n_2}})\varphi_{k_{n_1},k_{n_2}}(\theta_{j,k_{n_1},k_{n_2}},l_j)$ is smooth with respect to $k_x,k_y$. The fall out from the experimental realization of localization of the 5QW is to prove rigorously the following.

**Theorem IV.1:** $\lim_{t \to \infty} \sum_{j=2}^{5} \Psi(n_1, n_2, t; l; \alpha, \beta, \gamma, \lambda, \mu) = 0$ for $l = 1, 2, 3, 4, 5$ and $\alpha, \beta, \gamma, \lambda, \mu \in C$ with $|\alpha|^2 + |\beta|^2 + |\gamma|^2 + |\lambda|^2 + |\mu|^2 = 1$.

**Proof:** Let us begin by noting in matrix form we can write $\sum_{j=2}^{5} \begin{bmatrix} \Psi(n_1, n_2, t; 1; \alpha, \beta, \gamma, \lambda, \mu) \\ \Psi(n_1, n_2, t; 2; \alpha, \beta, \gamma, \lambda, \mu) \\ \Psi(n_1, n_2, t; 3; \alpha, \beta, \gamma, \lambda, \mu) \\ \Psi(n_1, n_2, t; 4; \alpha, \beta, \gamma, \lambda, \mu) \\ \Psi(n_1, n_2, t; 5; \alpha, \beta, \gamma, \lambda, \mu) \end{bmatrix} = M \begin{bmatrix} \alpha \\ \beta \\ \gamma \\ \lambda \\ \mu \end{bmatrix}$, where $M = (m_{ij})_{1 \leq i, j \leq 5}$ is a $5 \times 5$ matrix whose entries are linear combinations of the following integrals

$$J_{n_1, n_2, t} = \frac{1}{4\pi^2} \int_{-\pi}^{\pi} \int_{-\pi}^{\pi} \frac{\cos(k_{n_1} n_1 + k_{n_2} n_2)}{\left[ 2\sqrt{2} D^{\frac{1}{2}}(k_{n_1}, k_{n_2}) - 2 \right] \cos \theta_{k_{n_1}, k_{n_2}}^{\pm, \pm}} \cos(\theta_{k_{n_1}, k_{n_2}} t) \, dk_{n_1} \, dk_{n_2} \text{ and}$$

$$K_{n_1, n_2, t} = \frac{1}{4\pi^2} \int_{-\pi}^{\pi} \int_{-\pi}^{\pi} \frac{\cos(k_{n_1} n_1 + k_{n_2} n_2)}{10\sqrt{2} \sin \theta_{k_{n_1}, k_{n_2}}^{\pm, \pm}} \sin(\theta_{k_{n_1}, k_{n_2}} t) \, dk_{n_1} \, dk_{n_2}.$$ It follows from the Riemann-Lebesgue

Lemma that for large times the integrals of any linear combinations of $J_{n_1, n_2, t}$ and $K_{n_1, n_2, t}$ become zero, and the proof is finished.

### V. Concluding Remarks

In this paper we have answered the localization problem for the 5QW. It is an interesting problem to study the localization problem for a general $M$ − particle quantum walk on a general $D$ − dimensional lattice in which the particles are allowed to stay at the same position, in addition to their original degrees of, $2^D$.